\documentclass{pas}

\usepackage{enumitem}
\usepackage{multirow}
\usepackage{amsmath}
\usepackage{multicol}
\usepackage{float}

\newcommand{\newvar}[2]{\expandafter\def\csname#1\endcsname{#2}}
\newcommand{\var}[1]{\csname#1\endcsname}

\newvar{1839}{GPM\,J1839\ensuremath{-}10}
\newvar{0715}{GPM\,J0715\ensuremath{-}1141}
\newvar{1627}{GLEAM-X\,J1627\ensuremath{-}52}
\newvar{1832 long}{ASKAP/DART\,J1832\ensuremath{-}0911}
\newvar{1832}{J1832\ensuremath{-}09}
\newvar{1813}{ZTF\,J1813\ensuremath{+}4251}
\newvar{1832p0u}{$P=2656.2412\pm0.0003$\,s}
\newvar{1832p1}{$P_1=2656.24978$\,s}
\newvar{1832p1u}{$P_1=2656.24978\pm0.00034$\,s}
\newvar{1832p2}{$P_2=2656.25325$\,s}
\newvar{1832p2u}{$P_2=2656.25325\pm0.00015$\,s}
\newvar{arsco}{AR\,Sco}
\newvar{1912}{J191213.72\ensuremath{-}441045.1}
\newvar{0715}{J0715\ensuremath{-}11}
\newvar{1810}{XTE\,J1810\ensuremath{-}197}
\newvar{rm alpha slope}{$m_\alpha = -0.418\pm0.012$\,m$^2$\,rad$^{-1}$}
\newvar{rm alpha offset}{$\rm{RM}_0=91.417\pm0.093$\,rad\,m$^{-2}$}
\begin{document}

\lefttitle{Publications of the Astronomical Society of Australia}
\righttitle{Csan\'ad Horv\'ath}

\jnlPage{1}{4}
\jnlDoiYr{2026}
\doival{10.1017/pasa.xxxx.xx}

\articletitt{Research Paper}

\title{Evidence for the binary nature of the long-period radio transient \var{1832 long}}

\author{\gn{Csan\'ad} \sn{Horv\'ath}$^{1,2}$, \gn{Natasha} \sn{Hurley-Walker}$^{1}$, \gn{Ziteng} \sn{Wang}$^{1}$, \gn{Andrew} \sn{Zic}$^{2}$, \gn{Emil} \sn{Lenc}$^2$}

\affil{$^1$International Centre for Radio Astronomy Research, Curtin University, Kent St, Bentley WA 6102, Australia\newline$^2$Australia Telescope National Facility, CSIRO, Space \& Astronomy, Epping, New South Wales, Australia}

\corresp{Csan\'ad Horv\'ath, \newline Email: csanadhorvath2002@gmail.com / csanad.horvath@postgrad.curtin.edu.au}

\citeauth{Horv\'ath, C., Hurley-Walker, N., Wang, Z., Zic, A. Evidence for the binary nature of \var{1832 long}. {\it Publications of the Astronomical Society of Australia} {\bf 00}, 1--12. https://doi.org/10.1017/pasa.xxxx.xx}

\history{(Received xx xx xxxx; revised xx xx xxxx; accepted xx xx xxxx)}

\begin{abstract}
Long-period transients are a class of periodic pulsed radio source repeating on the minute to hour timescale. Recently, an increasing number of them are being identified as binary systems, specifically white dwarfs with low-mass main-sequence companions. In this work we analyse the most luminous long-period transient discovered to date, \var{1832 long}, with two years of radio data, and propose that it, too, may be a white dwarf system, although in a far more compact orbit than the aforementioned.
The pulses are composed of quasi-periodic components which evolve in a systematic way over days and months. The source is highly linearly or elliptically polarised and its brightness enabled very high signal-to-noise measurement of the time-resolved Faraday rotation measure, which was found to vary across pulse phase. The linear polarisation position angle, circular polarised fraction, and spectral index also varied systematically in ways not typical of pulsars and magnetars.
We show that an ultra-compact asynchronous polar explains much of the phenomenology of \var{1832 long}, in particular the evolution of the pulse morphology, rotation measure variation, and periodic X-ray emission, although we cannot conclusively prove a binary nature. However, our model makes testable predictions.
\end{abstract}


\maketitle

\section{Introduction}

%%%%%%%%%% INTRO TO LPTS %%%%%%%%%%
Recent years have seen the discovery of an increasing number of repeating Galactic radio pulses with tens-of-minutes to hours-long periodicities, becoming known as long-period transients (LPTs). ``LPT'' --- also referred to as Long Period Pulsars (LPPs), Ultra-long period Pulsars (ULPs), Ultra-Long Period Magnetars (ULPMs), etc. --- is not yet a well-defined class of physical objects, but rather a phenomenological class of radio-sources which do not fit any other class, and might be heterogeneous in terms of origin. Their luminosity and highly polarised nature imply a coherent emission mechanism originating in a magnetic environment. Canonical theories of rotation-powered pulsars \citep{1975ApJ...196...51R} prohibits radio emission from neutron stars spinning so slowly. Their measured period derivatives (often upper limits) place almost all LPTs in or below the so-called pulsar death valley \citep{1993ApJ...402..264C, 2000ApJ...531L.135Z}. The dominant competing hypotheses for the origin of LPTs are white dwarfs with low-mass companions \citep{2024ApJ...976L..21H,2025ApJ...981...34Q,2025NatAs...9..672D, 2025A&A...695L...8R,2026ApJ...997..124Y,2026NatAs.tmp...27H,Rose2026} and magnetars slowed by fallback accretion \citep{2022ApJ...934..184R, 2022MNRAS.513L..68G, 2024ApJ...967...24F, 2024ApJ...970....2X}, among others \citep{2005ApJ...631L.143Z,2024PhRvD.109f3004B,2024ApJ...972...60X,2025A&A...704A.123N,2025ApJ...986...98Z,2025ApJ...988L..11M,2026JHEAp..5300593M}. \cite{2026JHEAp..5200566R} present a comprehensive review of the thus-far discovered LPTs and their theoretical interpretations.

%%%%%%%%%% CASE FOR BINARY LPTS %%%%%%%%%%
% Sentence here that x LPTs have binary nature confirmed. Frame section better
The progenitors of only three LPTs have been positively identified, based on optical counterparts: ILT\,J1101$+$55 \citep{2025NatAs...9..672D}, GLEAM$-$X\,J0704$-$37 \citep{2024ApJ...976L..21H, 2025A&A...695L...8R}, and ASKAP\,J174508.9\ensuremath{-}505149 \citep{Rose2026}. They are all binary systems comprising a white dwarf and a low-mass M-type companion star.
AR\,Scorpii \citep{2016Natur.537..374M,2017NatAs...1E..29B} and a similar white dwarf -- M-dwarf system \var{1912} \citep{2023NatAs...7..931P, 2024MNRAS.527.3826P, 2023A&A...674L...9S}, %and SDSS\,J230641.47+244055.8 \citep{2025MNRAS.tmp.1459C},
have also been observed to produce radio emission modulated by both the orbit ($\sim4$\,hours) and white dwarf spin period (1--5\,minutes). They are referred to in the literature as AR\,Sco - types or white dwarf pulsars.

Based on an 8.75\,hour periodic modulation of the radio pulses, \cite{2026NatAs.tmp...27H} found \var{1839} \citep{2023Natur.619..487H} to also have a binary nature, and found that the peculiar modulation pattern can be modelled in the same geometric framework as AR\,Scorpii and \var{1912}, suggesting analogous physical scenarios. In this model, the intensity of the radio emission and the arrival time of pulses is modulated by the relative alignment of the white dwarf's magnetic moment with the companion star, hence the detection of a pulse depends on both the white dwarf's spin phase and the orbital phase. Such features in an LPT might point towards a binary interpretation even where direct evidence at other wavelengths does not exist.
The emission mechanism itself is uncertain, but one candidate is relativistic electron-cyclotron maser emission \citep{2025ApJ...981...34Q,2026ApJ...997..124Y}.
There are only $\mathcal{O}(10)$ known LPTs and their distance, long periods, and often transient nature makes them difficult to study without long-term monitoring.

%%%%%%%%%% ITRO TO J1832-09 %%%%%%%%%%
\var{1832 long} (hereafter \var{1832}) is an LPT discovered contemporaneously by the Australian SKA Pathfinder (ASKAP) and the Daocheng Radio Telescope (DART) \citep{2025Natur.642..583W, 2024arXiv241115739L}. Its $\sim3$\,minute-long radio-pulses repeat every 44\,minutes reaching luminosities of up to $\sim10^{32}$\,erg~s$^{-1}$, making it the the most luminous LPT to date. It stands out as one of the few LPTs detected in X-rays  \citep[ASKAP\,J144834\ensuremath{-}685644 and ASKAP\,J174508.9\ensuremath{-}505149 are the others;][]{2025MNRAS.542.1208A, Rose2026}, reaching $\sim10^{33}$\,erg~s$^{-1}$, with the X-ray luminosity modulated on the radio period. It is at a dispersion-measure-inferred distance of $4.5^{+1.2}_{-0.5}$\,kpc. No infra-red or optical counterpart has been identified, although the firm detection of such is hampered by significant extinction at its distance and location in the Galactic plane.

%%%%%%%%%% DESCRIPTION OF PAPER %%%%%%%%%%
In this work we analyse a set of ASKAP observations at $\sim$800--1100\,MHz and 10\,s time resolution between December~2023 and September~2025.
The earliest detection of \var{1832} was in late 2023, reaching peak brightness in early 2024, and it has since faded by 3 orders of magnitude.
We describe the data and the derived pulse morphology and polarisation information in \autoref{sec:results}, and interpret the results in the geometric frame work of \cite{2026NatAs.tmp...27H} in \autoref{sec:interpretation}, proposing that \var{1832} is an ultra-compact asynchronous polar. It would have the tightest orbit of the binary LPTs and be a remarkable member of the cataclysmic variables.

\section{Data}\label{sec:data}

The Australian SKA Pathfinder \citep[ASKAP;][]{2021PASA...38....9H} is a 36-dish radio interferometer operating between 700--1800\,MHz in Inyarrimanha Ilgari Bundara, the CSIRO Murchison Radioastronomy Observatory. Equipped with phased array feeds (PAFs) that expand its field-of-view to 31\,deg$^2$ at 800\,MHz, it has been used for LPT discovery and follow-up, including J1832$-09$. Of particular relevance to this paper is its three-axis mount, including azimuth, elevation, and a third roll axis. This ensures that the PAFs remain oriented at a fixed angle to celestial sources, reducing distortion and simplifying polarisation calibration.
The primary goal of the instrument is a series of ongoing surveys covering science topics across cosmology, extragalactic astronomy, \textsc{Hi}, and transient radio sources \footnote{\href{http://atnf.csiro.au/projects/science/wide-area-surveys/askap-survey-science-projects/}{http://atnf.csiro.au/projects/science/wide-area-surveys/askap-survey-science-projects/}}. 

The Variables And Slow Transients survey \citep[VAST;][]{2021PASA...38...54M} and the Commensal Real-time ASKAP Fast Transients survey \citep[CRAFT;][]{2010PASA...27..272M, 2025PASA...42....5W} are two ASKAP surveys searching for radio transients. \var{1832} was discovered in the ASKAP observation conducted as part of these surveys through the detection of its bright, highly circularly polarised radio bursts with narrow subpulse structures \citep{2025Natur.642..583W}.
Follow-up observations revealed repeat emission and confirmed its classification as a long-period radio transient. These results motivated dedicated monitoring efforts, including target-of-opportunity observations and regular monitoring through VAST, to investigate its periodicity, variability, and polarimetric behaviour.

Project AS321 was undertaken across observing semesters 2025-APR and 2025-OCT to monitor \var{1832} and two other LPTs in the same field. The goal was consistent monitoring of these sources to determine the long-term evolution of their brightnesses and polarisation properties, leveraging ASKAP's wide field-of-view and polarisation purity. The data are described in \autoref{tab:observations}; the inconsistency in the monitoring cadence is largely due to conflicts with the large survey projects which take priority for the observatory. Nevertheless, in conjunction with the early discovery and follow-up data, the data represent one of the largest and most consistent monitoring programmes towards \var{1832}.

% Mention that they are all at the same frequency

\begin{table}[t]
    \small
    \centering
    \begin{tabular}{ccccc}
        \hline
        ID      & Beams      & Start / MJD    & Length / s & Band / MHz \\
        \hline
        SB55237 & 19         & 60286.287496 &   557 &  744--1031 \\
        SB58387 & 13         & 60341.117946 &  7206 &  800--1087 \\
        SB58609 & 13         & 60343.880811 & 28795 &  800--1087 \\
        SB58753 & 13         & 60346.874845 & 28865 &  800--1087 \\
        SB60091 & 13         & 60384.763558 & 30596 &  800--1087 \\
        SB64280 & 27         & 60521.398205 & 32408 &  800--1087 \\
        SB64328 & 13         & 60524.396548 & 32398 &  800--1087 \\
        SB64345 & 19         & 60525.376393 & 28805 &  800--1087 \\
        SB73422 & 02         & 60797.668154 &  1802 &  744--1031 \\
        SB73824 & 19         & 60811.644008 &  1802 & 1296--1439 \\
        SB74078 & 00, 02, 15 & 60814.868546 &  1861 &  744--1031 \\
        SB74155 & 19         & 60816.749883 &  1861 & 1296--1439 \\
        SB74207 & 14, 19, 20 & 60821.855883 &  1802 &  800--1087 \\
        SB74356 & 02, 15     & 60835.563087 &  1802 &  744--1031 \\
        SB74438 & 14, 19, 20 & 60839.594283 &  1802 &  800--1087 \\
        SB74453 & 14, 19, 20 & 60840.669323 &  1802 &  800--1087 \\
        SB75009 & 00, 02, 15 & 60880.447855 &  1802 &  744--1031 \\
        SB75057 & 14, 19, 20 & 60881.523605 &  1802 &  800--1087 \\
        SB75072 & 14, 19, 20 & 60882.597908 &  1802 &  800--1087 \\
        SB75087 & 14, 19, 20 & 60883.672771 &  1802 &  800--1087 \\
        SB75096 & 14, 19, 20 & 60884.479008 &  1802 &  800--1087 \\
        SB75103 & 14, 19, 20 & 60884.747884 &  1802 &  800--1087 \\
        SB76741 & 00, 02, 15 & 60929.353606 &  2449 &  744--1031 \\
        SB76766 & 00, 02, 15 & 60929.626975 &  1802 &  744--1031 \\
        SB76845 & 00, 02, 15 & 60932.577897 &  2359 &  744--1031 \\
        SB76903 & 00, 02, 15 & 60933.656474 &  2160 &  744--1031 \\
        SB76973 & 00, 02, 15 & 60936.613028 &  2249 &  744--1031 \\
        SB76982 & 00, 02, 15 & 60937.421072 &  1802 &  744--1031 \\
        SB77009 & 14, 19, 20 & 60939.328402 &  1802 &  800--1087 \\
        SB77017 & 00, 02, 15 & 60939.572538 &  1802 &  744--1031 \\
        SB77046 & 14, 19, 20 & 60940.424748 &  1802 &  800--1087 \\
        SB77135 & 00, 02, 15 & 60943.333455 &  1802 &  744--1031 \\
        SB77179 & 00, 02, 15 & 60944.408506 &  1802 &  744--1031 \\
        SB77222 & 00, 02, 15 & 60945.483998 &  1811 &  744--1031 \\
        SB77285 & 14, 19, 20 & 60947.633108 &   896 &  800--1087 \\
        \hline
    \end{tabular}
    \caption{Table of ASKAP observations used in this work. \label{tab:observations}}
\end{table}

\section{Results}\label{sec:results}

\begin{figure*}[p]
    \centering
    \includegraphics[width=0.87\linewidth]{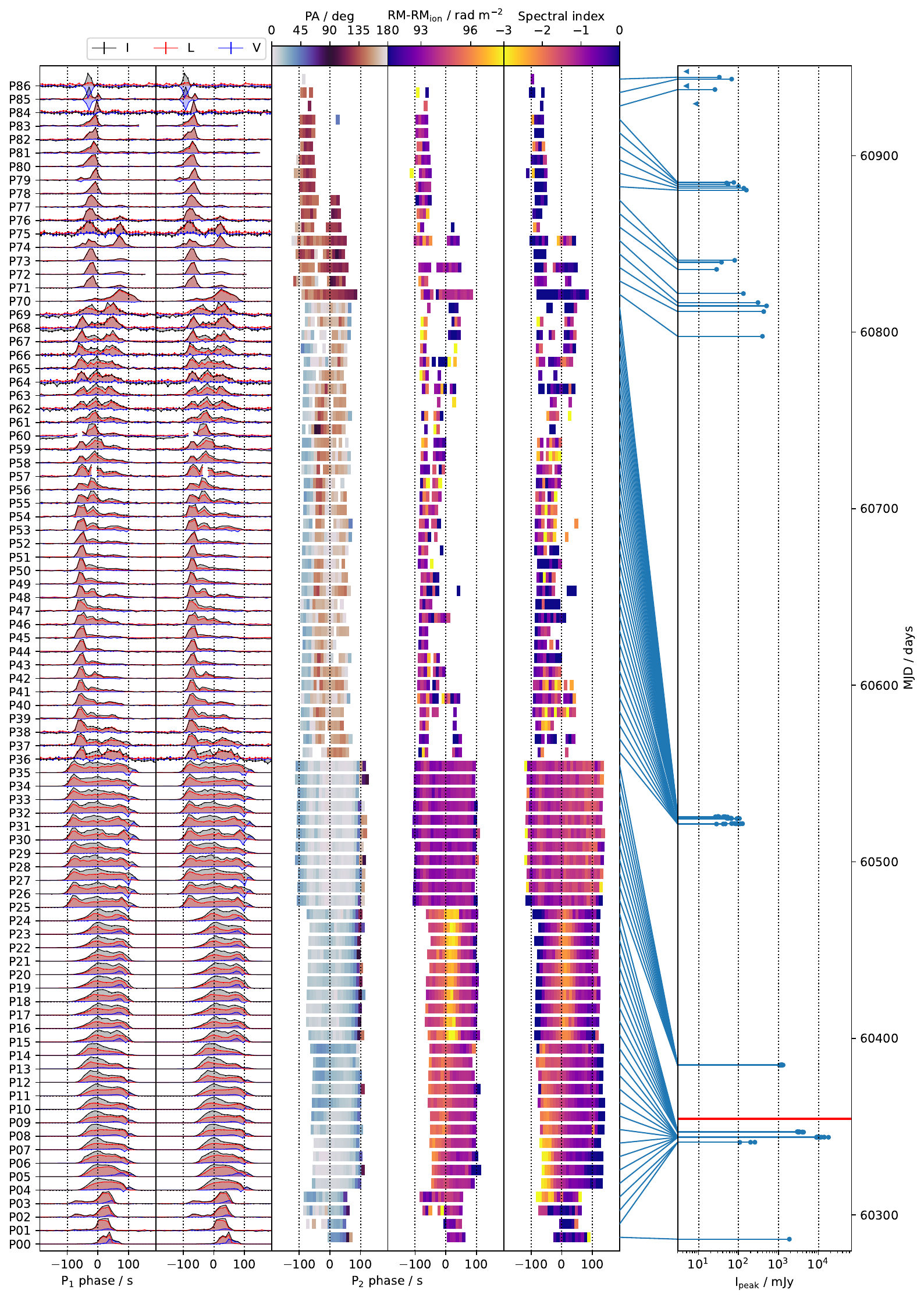}
    \caption{\var{1832} pulses detected by ASKAP in chronological order (bottom to top), with the pulse name and topocentric UTC times of pulse centres marked at left. The leftmost column is folded on \var{1832p1}; the rest are folded on \var{1832p2}. The columns from left to right are: peak-normalised mean pulse profile over the band with linear and circular polarised components; linear polarisation position angle (PA); Faraday rotation measure (RM) minus the ionospheric contribution; spectral index; peak mean intensity across the band ($I_{\rm{peak}}$) with absolute time on the vertical axis. The measurement of these values is described in \autoref{subsec:dynspec_model}. The horizontal red line indicates the 2024-02-14 Chandra X-ray observation \citep{2025Natur.642..583W}. The PA and circular fraction are shown with greater clarity in \autoref{fig:1832_pa_vfrac_phase}.}
    \label{fig:1832_pulsestack}
\end{figure*}

%%%%%%%%%% STRUCTURE OF RESULTS %%%%%%%%%%
This section details the observed properties of the radio emission of \var{1832}, organised into pulse morphology, quasi-periodicity, linear polarisation, circular polarisation, and spectral shape. \autoref{fig:1832_pulsestack} aggregates the results. The methods used are summarised here and details not pertinent to understanding the results are described in the Appendix. Physical interpretations are made in \autoref{sec:interpretation}.

\subsection{Pulse morphology}

%%%%%%%%%% PULSE STACK AND OLD PERIODS %%%%%%%%%%
We calculated light curves for each pulse in Stokes I, L, and V by averaging the dynamic spectrum over frequency, presented in \autoref{fig:1832_pulsestack}. A pulse was detected at 87 of the 91 predicted times of arrival, with 4 non-detections in September 2025 shown as flux density upper limits in the right-most panel.
The published period \var{1832p0u} \citep{2025Natur.642..583W} was found to to be inaccurate over the 2 years of data. 
In the leftmost panel, the pulses are stacked on the period \var{1832p1u} which best aligns the pulse centroids, measured by fitting a single Gaussian to each pulse light curve. % (see \autoref{fig:1832_single_fit} and \autoref{fig:1832_single_period_fit}).
The pulses are labelled chronologically as P00, P01, P02, etc. Indices separated by a dash indicate a range of pulses, e.g. P01--03 refers to P01, P02, and P03 collectively.

\begin{figure}[ht]
    \centering
    \includegraphics[width=1.0\linewidth]{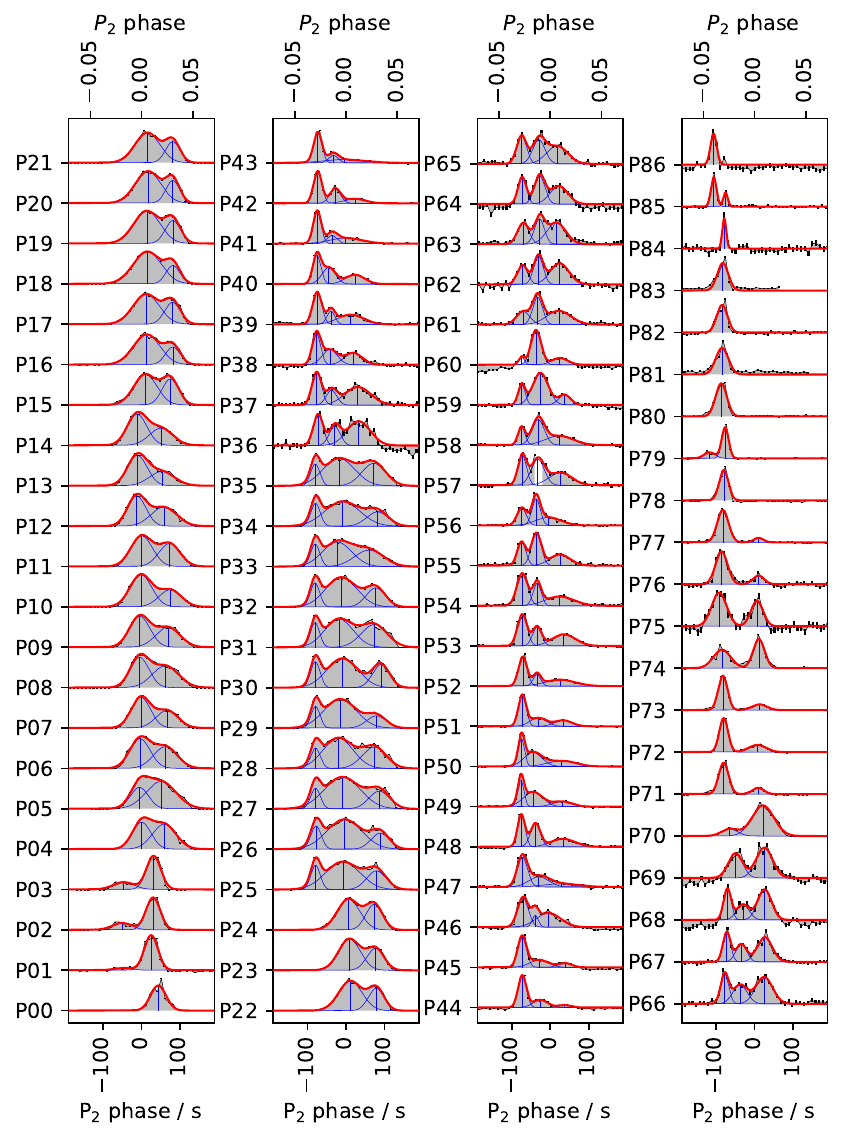}
    \caption{\var{1832} peak-normalised mean pulse profiles over the ASKAP band in chronological order (bottom to top, left to right), with the pulse name marked at left. The pulses are folded on \var{1832p2}. Each pulse has been modelled as a sum of Gaussian components drawn in blue, with the resulting modelled pulse drawn in red.}
    \label{fig:1832_askap_multi_gauss}
\end{figure}

%%%%%%%%%% SUBSTRUCTURE MODEL %%%%%%%%%%
Most pulses, however, are composed of two or three sub-pulse components, the width and spacing of which is often consistent between consecutive pulses. Many non-consecutive pulses separated by days or weeks seem to be composed of components with the same distribution of pulse phases, sometimes with the absence or addition of components (e.g. P24--25, P35--36, P47--48).
We modelled each light curve $S$ as the sum of $m$ components as
\begin{equation}
    S'(\phi) = \sum_{j=1}^{m} A_{j} \exp\left(-\frac{(\phi-\mu_{j})^2}{2\sigma_{j}^2}\right)
\end{equation}
where $S'(\phi)$ is the modelled light curve, $\phi$ is the period phase, $\mu_{j}$ is the component centroid phase, $A_{j}$ is the component amplitude, and $\sigma_{j}$ is the component width. The fit for each pulse was done using \verb|scipy.optimize.curve_fit|, initialised with width $\sigma_{j} = 5$\,s and the following $\mu_{j}$ centroids:
\setlength{\columnsep}{0pt}
\begin{multicols}{2}\begin{description}
    \item[         P00:] 40s                  
    \item[     P01--03:] -66s, 28s        
    \item[     P04--24:] 0s, 45s          
    \item[     P25--35:] -80s, 0s, 75s    
    \item[     P36--68:] -80s, -20s, 20s  
    \item[     P69--77:] -75s, 0s         
    \item[P78, P80--84:] -75s        
    \item[P79, P85--86:] -100s, -75s 
\end{description}\end{multicols}
The amplitudes $A_{j}$ were initialised as the value of the pulse profile at the initial centroid. To avoid over-fitting with too many components, the Bayesian Information Criterion (BIC) was approximated \citep{Kass01061995}. BIC rewards closeness-of-fit but penalises too many parameters. It is calculated as
\begin{equation}
    \text{BIC} \approx \chi^2 + 3m  \ln n ~~\text{where}~~ \chi^2 = \sum_{i=1}^n \left(\frac{S_i-S'(\phi_i)}{\Delta S_i}\right)^2 \text{,}
\end{equation}
where $S_i$ and $\Delta S_i$ are the pulse amplitude and associated uncertainty at time-step $i$, $n$ is the number of time-steps, and $3m$ is the number of parameters.
If the removal of a component reduced the BIC, that component was removed and the fit repeated. The modelled profiles are presented in \autoref{fig:1832_askap_multi_gauss}.

\begin{figure}[ht]
    \centering
    \includegraphics[width=1.0\linewidth]{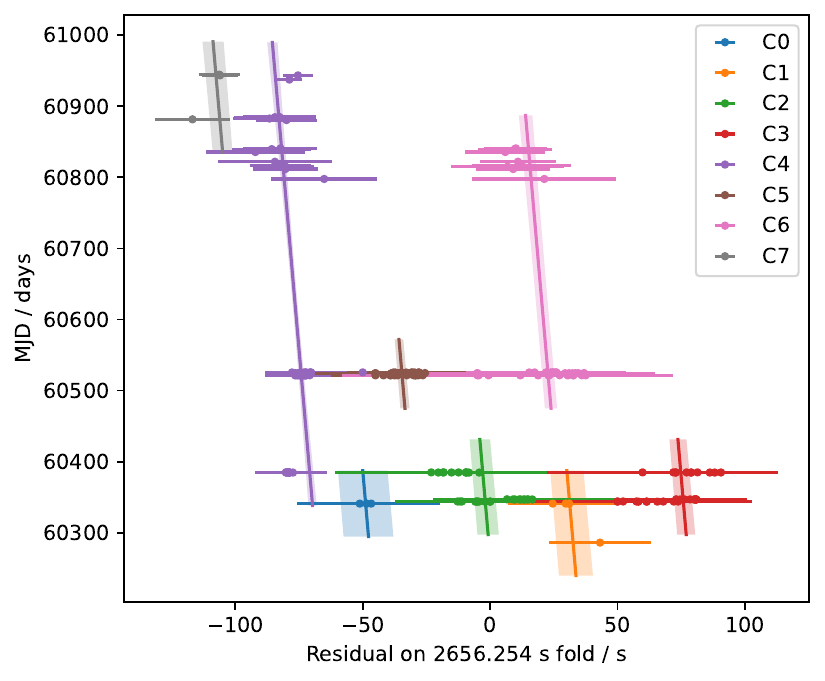}
    \caption{Linear fit to the phases (folded on 2656.254\,s) of the \var{1832} Gaussian component centroids. The best-fit linear model $\mu_k = m (t-60615 \text{MJD}) + c_k$ is shown with $m = (-2.83\pm0.57)\times10^{-7}$\,s\,s$^{-1}$ and $(c_7, c_4, c_0, c_5, c_2, c_6, c_1, c_3) = (-102.4\pm3.9, -78.7\pm1.0, -57.9\pm10.0, -39.0\pm1.7, -10.6\pm4.0, 18.3\pm2.4, 22.3\pm6.5, 67.1\pm3.3)$\,s.}
    \label{fig:1832-09_period_fit}
\end{figure}

%%%%%%%%%% SUBSTRUCTURE ALIGNING PERIOD %%%%%%%%%%
Next, we found the period which best aligns the substructure component centroids. In the absence of (near) continuous data, the criteria for whether a particular pulse component corresponds with a component of a different pulse depends on human judgement, so we grouped the components which we deemed to align with each other when folded on an initial guess period of 2656.254\,s based on the clustering of the centroid residuals in \autoref{fig:1832-09_period_fit} and visual similarity in ambiguous cases. We labelled the components as \begin{verbatim}
P[pulse index].[left-to-right index within pulse]    
\end{verbatim} and labelled the groups as C0 to C7. The groups are composed of the following pulse components: 
\setlength{\columnsep}{0pt}
\begin{multicols}{2}\begin{description}    
    \item[C0:] P01--03.0                   
    \item[C1:] P00.0, P01--03.1            
    \item[C2:] P04--24.0, P25--35.1        
    \item[C3:] P04--24.1, P25--35.2        
    \item[C4:] P25--78.0, P80--84.0, P85.1 
    \item[C5:] P36--68.1                   
    \item[C6:] P36--68.2, P69--77.1        
    \item[C7:] P79.0, P85--86.0            
\end{description}\end{multicols}

We fit a linear model $\mu_k = m (t-60615 \text{MJD}) + c_k$ with slope $m$ and individual time-intercepts $c_k$ for each component group $k\in\{0,1,...,7\}$ to the pulse component centroid phases $\mu_k$ (when folded on 2656.254\,s). The fit was done using \verb|scipy.optimize.curve_fit|, with the component width parameter $\sigma$ used as the data point uncertainty. The resulting fit is presented in \autoref{fig:1832-09_period_fit}. We find a best-fit slope $m = (-2.83\pm0.57)\times10^{-7}$\,s\,s$^{-1}$ corresponding to a period \var{1832p2u}, which is the stacking period in the second panel onwards of \autoref{fig:1832_pulsestack}.

The best-fit time-intercepts, in ascending order, were $(c_7, c_4, c_0, c_5, c_2, c_6, c_1, c_3) = (-102.4\pm3.9, -78.7\pm1.0, -57.9\pm10.0, -39.0\pm1.7, -10.6\pm4.0, 18.3\pm2.4, 22.3\pm6.5, 67.1\pm3.3)$\,s.
P00--03 appear misaligned from the rest by an offset of $\sim-50$\,s, despite being aligned with each other and having similar peak-to-peak spacing as P04--35. The Stokes V profile and PA sweep at the end of the P00--03 pulses also appears similar to, but offset from, the subsequent pulses, implying a stronger association between the components of P00--03 and P04--35 than just similarity in the Stokes I profile. It is possible that some components wander in pulse phase (for example, the first component of P05--14 appears to drift to the left). Alternatively, C1 and C6 might actually belong to the same group since $c_1 \approx c_6$, with that pulse component temporarily disappearing in P04--35. Since this is up to interpretation, we only grouped components which could be confidently associated.

\subsection{Quasi-periodicity}

\begin{figure}[t]
    \centering
    \includegraphics[width=1.0\linewidth]{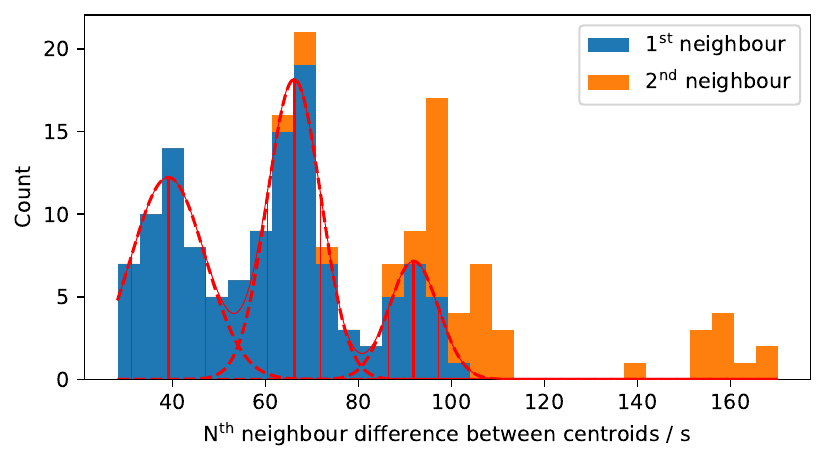}
    \caption{Histogram of the first and second neighbour differences between the centroids of Gaussian components of the \var{1832} pulses.}
    \label{fig:1832-09_quasi_period}
\end{figure}

\autoref{fig:1832-09_quasi_period} is a histogram of the time-differences between 1$^\text{st}$ and 2$^\text{nd}$ neighbour pulse component centroids. By modelling the 1$^\text{st}$ neighbour distribution as a sum of normal distributions, we find 3 peaks at $ 39 \pm 6$\,s, $ 65 \pm 7$\,s, and $ 92 \pm 5$\,s where the uncertainties are the $1\sigma$ widths of the components. The first 2$^\text{nd}$ neighbour peak coincides with the 92\,s peak, but neither of the 2$^\text{nd}$ neighbour peaks coincide with an integer multiple of any of the 1$^\text{st}$ neighbour peaks. This can be interpreted as follows: pulse components are preferentially centred at specific phases, maintaining their separation as other components appear and disappear, but the distribution of preferential phases is not strictly periodic.

\begin{figure}[t]
    \centering
    \includegraphics[width=1.0\linewidth]{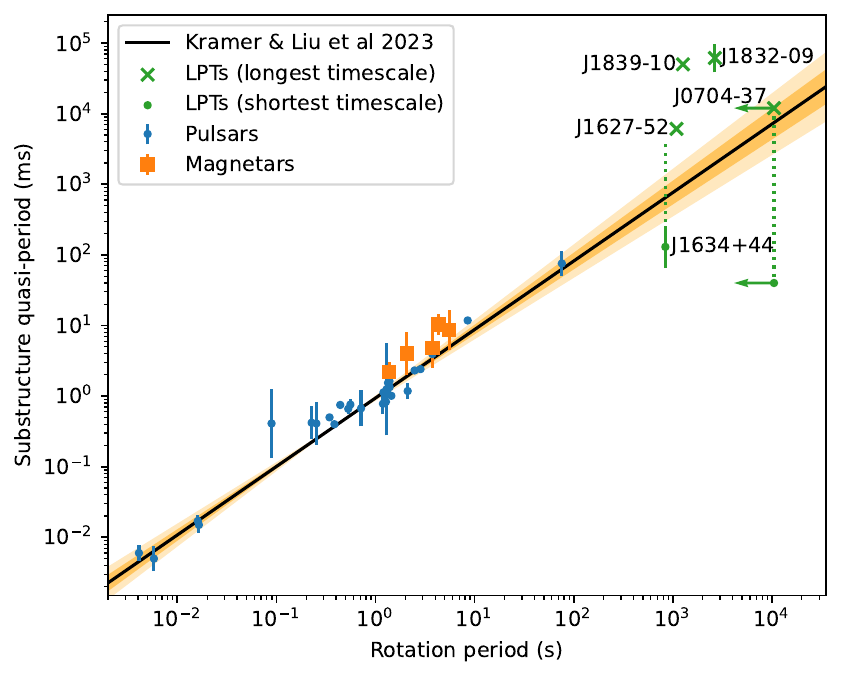}
    \caption{Pulsar and magnetar quasi-periods and $P_\mu = (1.12\pm0.14)P^{1.03\pm0.04}$\,ms relationship from  from \cite{2024NatAs...8..230K} and LPT quasi-periods \citep{2025ApJ...988L..29D,2022Natur.601..526H,2023Natur.619..487H,2024ApJ...976L..21H}. The dark and light shaded regions represent the $1\sigma$ and $2\sigma$ confidence intervals from \citeauthor{2024NatAs...8..230K}'s fit to the pulsar and magnetar points. The short timescale quasi-periods of J1634+44 and J0704\ensuremath{-}37 are based on CHIME/Pulsar and PTUSE data respectively, which are not sensitive to slower variation on the timescale of the pulse width. The dotted line on J1634+44 extends to the upper bound of the longest timescale quasi-period given by the pulse width. The left arrow on J0704\ensuremath{-}37 represents the fact that only the orbital period is known and the white dwarf spin period may or may not be shorter.}
    \label{fig:period_quasi-period_scatter}
\end{figure}

A correlation between pulse substructure quasi-period and spin period has been proposed for pulsars and magnetars by \cite{2024NatAs...8..230K}. They find that the quasi-period follows $P_\mu = (1.12\pm0.14)P^{1.03\pm0.04}$\,ms, where $P$ is the spin period and the uncertainties are $1\sigma$ intervals. According to this, the predicted quasi-period of \var{1832} should be $P_\mu = 3.8\pm1.7$\,s, compared with the observed $P_\mu \sim 65$\,s. Note that pulses also exhibit fine (0.5-s timescale) structure when observed at high time resolution with CRACO \citep{2025Natur.642..583W} which may also be quasi-periodic. The pulsar and magnetar quasi-periods used by \citeauthor{2024NatAs...8..230K} and the LPTs for which we found such a measurement in the literature are shown in \autoref{fig:period_quasi-period_scatter}. Interestingly, the \var{1832} quasi-period is quite close to that of \var{1839}. LPTs do not follow the trend set by neutron stars, although their number is still too small to draw firm conclusions.

\subsection{Linear polarisation}\label{subsec:linpol}

%%%%%%%%%% DESCRIPTION OF RM %%%%%%%%%%
\var{1832} is mostly linearly polarised, usually about 50\% to 90\%. Propagation through a magnetised plasma causes the plane of linear polarisation to rotate by an angle RM$\lambda^2$, where RM is the Faraday rotation measure. Most of the RM of Galactic sources is typically attributed the interstellar medium (ISM), but plasma local to the source, as well as the Earth's ionosphere contribute as well. The measured RM is described by the following sum:
\begin{equation}
    \text{RM}=\text{RM}_\text{src}+\text{RM}_\text{ISM}+\text{RM}_\text{ion}.
\end{equation}
For each time-bin, we measured the PA and RM by fitting a sinusoid to Stokes Q and U in $\lambda^2$ (see \autoref{subsec:dynspec_model}). Throughout this work, by PA we refer to the position angle $\psi_\nu$ at infinite frequency $\nu \rightarrow \infty$. Where there was insufficient signal-to-noise to measure the RM, we tried to measure the PA by fixing RM to $+90.5\pm0.1$\,rad\,m$^{-2}$ \citep{2025Natur.642..583W}. The third and fourth columns of \autoref{fig:1832_pulsestack} display these results where measurement was possible.

%%%%%%%%%% RM VARIATION %%%%%%%%%%
The average pulse-to-pulse RM was found to vary on a daily cadence in a way which was consistent with the variation in the Earth's known ionospheric RM contribution. We used the \verb|spinifex| \citep{2025ascl.soft07016M} package to access historical ionospheric data and subtracted RM$_\text{ion}$ from the measured RM. Aggregated results for 4 epochs of the data are presented in \autoref{tab:rm}. Subtracting RM$_\text{ion}$ reduced the standard deviations and brought the P25--36 mean (which shows minimal phase-resolved RM variation in \autoref{fig:1832_pulsestack}) into close agreement with the P36--86 mean. On the other hand, P04--14 and P15--24 have significant RM excess associated with specific pulse phases in \autoref{fig:1832_pulsestack}; the beginnings of P04--14 and the middles of P15--24.

%%%%%%%%%% SOME ANALYSIS %%%%%%%%%%
The shape of the RM curve in P04--24 appears consistent between consecutive pulses, but changes between non-consecutive pulses, implying systematic changes in the electron density and/or magnetic fields near the source itself rather than stochastic turbulence of the interstellar medium. Phase-resolved RM variations across the pulse have been reported in pulsars and magnetars \citep{2009MNRAS.396.1559N, 2019MNRAS.483.2778I, 2024NatAs...8..617D} and have been observed in repeating fast radio burst sources \citep{2023MNRAS.520.2039Y}, but these are not usually as structured as seen in \var{1832}.

\begin{figure}[ht]
    \centering
    \includegraphics[width=\linewidth]{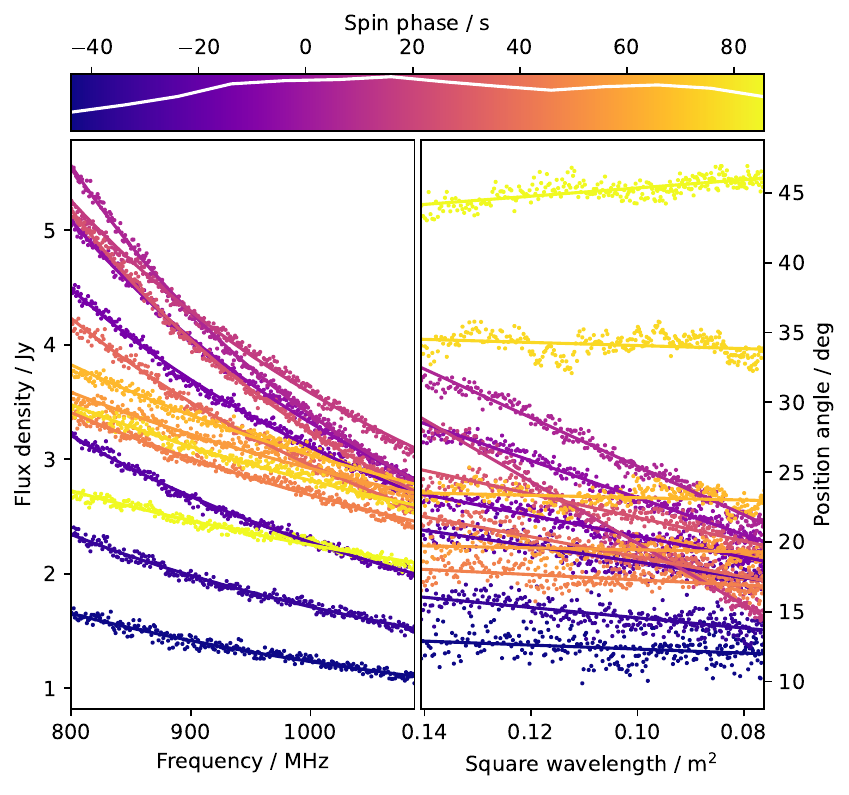}
    \caption{Spectrum and position angle of P18. Data is shown as dots and fits are shown as continuous lines. Colour represents pulse phase, defined by the colour bar at right, which includes the pulse profile in white.}
    \label{fig:rep_spec}
\end{figure}

Linear to circular polarisation conversion in the magnetosphere can result in apparent RM variation when generalised Faraday conversion is incorrectly modelled (see \cite{2024NatAs...8..606L} for an example and in-depth analysis), which should be correlated with increased circular polarisation and PA variation, both in a frequency-dependent way. While we do see circularly polarised features in \var{1832} (\autoref{fig:1832_pulsestack}, \autoref{fig:1832_pa_vfrac_phase}), these are not coincident with the RM features in spin phase, and occur in pulses both with and without RM variation. Furthermore, the PA maintains a linear relationship with $\lambda^2$  (\autoref{fig:rep_spec}), implying that the measured RM change is genuine rotation by cold, weakly magnetised plasma following the standard Faraday rotation law.

%%%%%%%%%% TABLE OF RM %%%%%%%%%%
\begin{table}[ht]
    \centering
    \begin{tabular}{ccccc}
        \hline
        Pulses  & \multicolumn{2}{c}{RM} & \multicolumn{2}{c}{RM$-$RM$_\text{ion}$} \\
        \hline
                & Mean       & STD       & Mean                & STD                \\
        P00--14 & 89.84      & 1.30      & 94.13               & 0.81               \\
        P15--24 & 91.61      & 1.61      & 95.12               & 1.42               \\
        P25--35 & 90.69      & 0.89      & 93.41               & 0.45               \\
        P36--86 & 91.78      & 1.26      & 93.42               & 1.10               \\
        \hline
    \end{tabular}
    \caption{Weighted mean and standard deviation of the phase-resolved RM with and without the ionospheric contribution subtracted. All units are rad\,m$^{-2}$. \label{tab:rm}}
\end{table}

%%%%%%%%%% POSITION ANGLE %%%%%%%%%%
The PA in P00--35 was a constant $\sim20^\circ$ for most of the pulse, except for a rapid sweep with positive gradient in the final 30\,seconds of those pulses, as seen in \autoref{fig:1832_pulsestack} and \autoref{fig:1832_pa_vfrac_phase}. P36--69 alternated between $\sim-50^\circ$ and $\sim20^\circ$. P70--86 began with a PA sweep with negative gradient before settling at $\sim-50^\circ$. The PA curve appears to evolve slowly over months and does not show the hallmark S-shaped curve associated with rotating dipole emitters.

\subsection{Circular polarisation}

\begin{figure}[ht]
    \centering
    \includegraphics[width=1.0\linewidth]{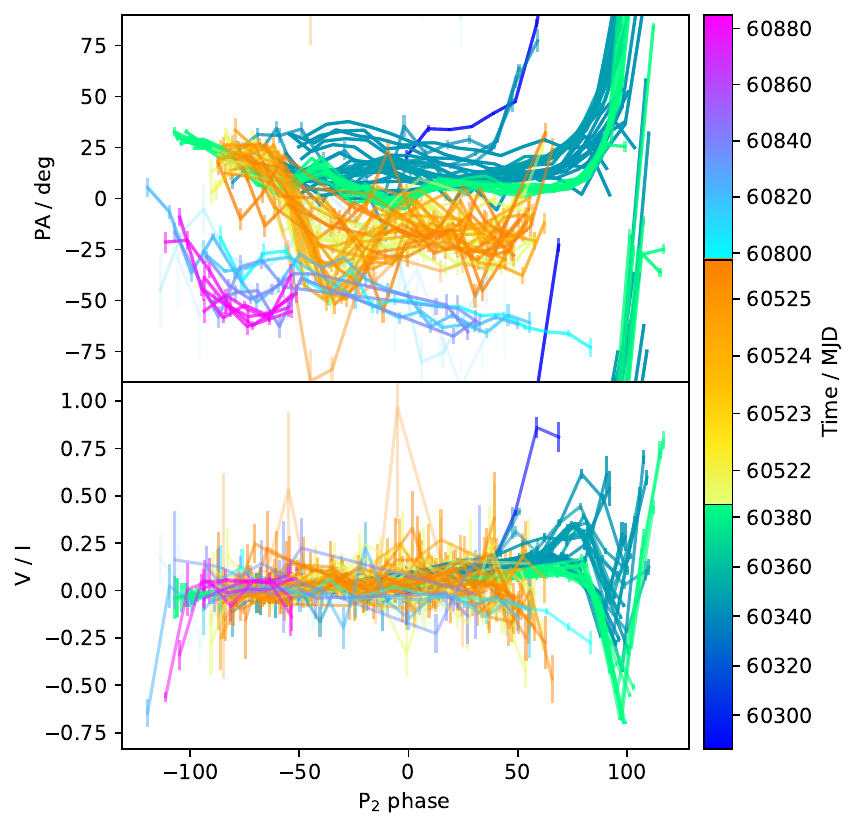}
    \caption{The linear polarisation position angle (PA) and circular polarised fraction (V/I) against \var{1832p2} phase. The colour scale shows the time of arrival, split into three epochs.}
    \label{fig:1832_pa_vfrac_phase}
\end{figure}

%%%%%%%%%% CIRCULAR PULSE COMPONENTS AND CORRELATION WITH PA %%%%%%%%%%
Circular polarisation was present at the ends of P00--35 and the starts of P79 and P85--86, sometimes exceeding 90\%. The handedness of the circularly polarised component alternates and it seems to switch at particular pulse phases in consecutive pulses, and even pulses separated by several weeks. \autoref{fig:1832_pa_vfrac_phase} shows the fractional circular polarisation and the PA over pulse phase for all pulses. The phases of increased circular polarised fraction coincide with the phases of PA sweep, shown most clearly at the ends of P00--35. Despite this connection, the handedness of the circular polarisation seems uncorrelated with the direction of PA sweep. 

P79 and P85--86 have a similar pattern to P00--35, but in reverse. These pulses start with a PA sweep and high circular polarisation and include a new pulse component at earlier phase than all other pulses. In general, it seems that the sub-pulse phase, circular polarised fraction, and linear PA are somehow connected. The presence of these features did not, however, coincide with the RM structure discussed earlier.

\subsection{Spectral shape}

%%%%%%%%%% POWER LAW %%%%%%%%%%
The Stokes I spectrum closely follows a power law with negative spectral index $\alpha_I$, shown for a representative pulse in \autoref{fig:rep_spec}, and for all pulses in the fifth column of \autoref{fig:1832_pulsestack}. We only probed a narrow slice of the spectrum, but \var{1832} has been seen up to 5\,GHz, while its spectral turnover renders it invisible at $\lesssim300$\,MHz \citep{2025Natur.642..583W}. We measured $\alpha_I$ by fitting a power law to each time-step of the de-dispersed dynamic spectrum (\autoref{subsec:dynspec_model}). Of the high signal-to-noise observations (P00--35), P25--35 had the most stable spectral index across pulse phase, with a mean of $-1.4$ and standard deviation of 0.3. Conversely, P04--14 and P15--24 varied significantly in a smooth, non-stochastic way. The rest of the pulses with much lower signal to noise had either apparently flat spectra or $\alpha_I\sim-1$.

\begin{figure}
    \centering
    \includegraphics[width=\linewidth]{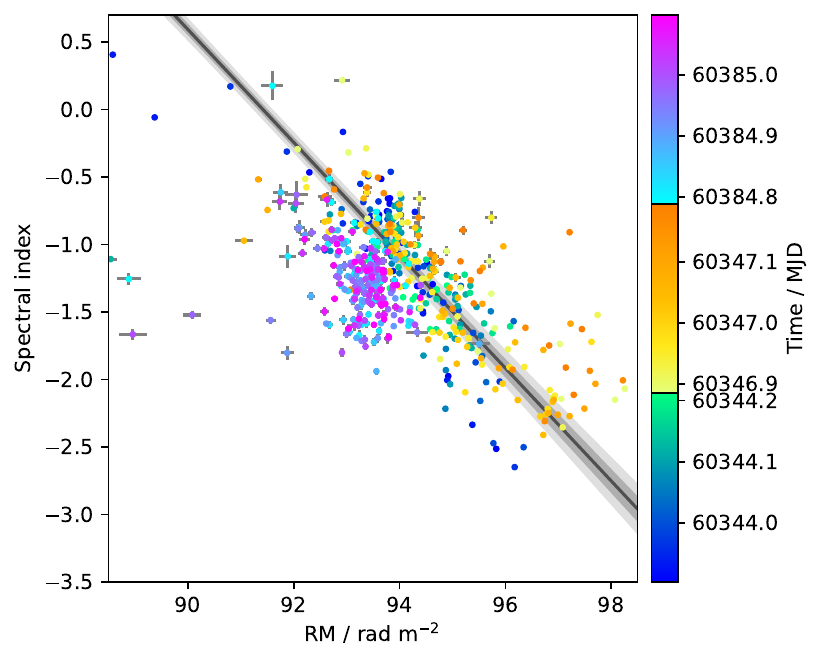}
    \caption{Spectral index and rotation measure of P04--35. Only time-bins with $L>10\sigma$ are included. The linear fit $\alpha=(\mathrm{RM}-\mathrm{RM}_0)m_\alpha$ has parameters \var{rm alpha slope} and \var{rm alpha offset}.}
    \label{fig:rm_alpha_scatter}
\end{figure}

%%%%%%%%%% CORRELATION WITH RM %%%%%%%%%%
In \autoref{fig:1832_pulsestack}, $\alpha_I$ appears correlated with RM, peaking at the same pulse phases in P03-P24. \autoref{fig:rm_alpha_scatter} shows RM against $\alpha_I$, making a negative correlation evident. Like the RM variation, this is not a fitting artefact; the spectrum maintains a power-law shape and the PA stays linear with $\lambda^2$ throughout the pulses. A model of the form $\alpha_I=(\mathrm{RM}-\mathrm{RM}_0)m_\alpha$ has best-fit parameters \var{rm alpha slope} and \var{rm alpha offset}.

\section{Binary interpretation}\label{sec:interpretation}

In this section we investigate the consistency of the geometric binary model from \cite{2026NatAs.tmp...27H} with the properties of our \var{1832} data as described in \autoref{sec:results}, and discuss the extent to which such a model has explanatory and predictive power for this system.
Based on  its pulse profile evolution and polarisation features, we suggest that \var{1832} is a magnetic white dwarf in a binary whose orbit and spin periods are slightly asynchronous. This scenario has already been proposed for the 1.16-hour LPT ASKAP\,J175534.9\ensuremath{-}252749.1 \citep{2025MNRAS.542..203M}, and a higher-integer near-resonance has been proposed for the 14-minute LPT CHIME/ILT\,J1634\ensuremath{+}4450  \citep{2025A&A...699A.341B}. Slight asynchronism, in our case, is motivated by the observation of slow changes in pulse morphology and polarisation properties associated with the gradual drift of the white dwarf's orientation with respect to the orbit. Slight asynchronism is also required by a number of proposed LPT emission mechanisms \citep{2025ApJ...981...34Q,2026ApJ...997..124Y,2026ApJ...999L...2Z}. All of these works predict luminosities several orders of magnitude below what was observed for \var{1832} (\autoref{sec:luminosity}), but we hesitate to over-interpret this since with a 44\,min orbit we are well outside of the regime for which those calculations were intended.

A 44-minute orbit necessitates that a main-sequence companion would fill its Roche-lobe and be stripped of its outer layers, possibly making \var{1832} akin to an AM Canum Venaticorum (AM CVn) star, but the spin-orbit near-synchronism implies a better classification may be an ultra-compact polar. There are a handful of known polars where the white dwarf spin and orbital periods are different by $\lesssim 1\%$, known as asynchronous polars \citep{2002AIPC..637....3W}, which are synchronised over centuries by torque from the white dwarf's strong magnetic field \citep{1983MNRAS.205.1031C}.
% A double white dwarf is also possible \citep{2026ApJ..1004L..46Z}, but we do not explore this further.
The subsequent subsections describe the model and presents arguments in favour of a binary interpretation of the data.

\subsection{Geometry} \label{subsec:geometry}

Following \cite{2026NatAs.tmp...27H}, the white dwarf is placed at the origin with $\hat{z}$ as the spin axis. The orbital axis is inclined about $\hat{y}$ by angle $i$, and the magnetic moment $\mu$ is inclined from $\hat{z}$ by angle $\alpha$. The observed pulse amplitude depends on the angle $\beta = \arccos(\hat{\mu}\cdot\hat{r}_\text{MD})$ between the magnetic moment $\hat{\mu} = (\cos\phi\sin\alpha, ~\sin\phi\sin\alpha, ~\cos\alpha)$ and the line of sight $\hat{r}_\text{E} = (\sin\zeta\cos\phi_0, ~\sin\zeta\sin\phi_0, ~\cos\zeta)$, and the angle $\beta_\text{MD} = \arccos(\hat{\mu}\cdot\hat{r}_\text{MD})$ between $\hat{\mu}$ and the companion at $\hat{r}_\text{MD} = (\cos{i}\cos\phi_\text{orb}, ~\sin\phi_\text{orb}, -\sin{i}\cos\phi_\text{orb})$. Here, $\phi$ is the white dwarf spin phase, $\phi_\text{orb}$ is the orbital phase, and $(\phi_0,\zeta)$ defines the line of sight in spherical coordinates. The structure of the radio beam is described by
\begin{equation}\label{eqn:beam}
    f_\text{spin}(\beta) = \exp \left( \frac{-\beta^2}{2(W_\text{spin}/5)^2} \right) \cos\left(\beta\theta_\text{struc}\right),
\end{equation}
where $W_\text{spin}$ sets the beam width. This differs from \cite{2026NatAs.tmp...27H} by the addition of the $\cos\left(\beta\theta_\text{struc}\right)$ factor, which creates a beam composed of concentric cones whose radial periodicity is set by $\theta_\text{struc}$. The actual beam might not be structured this way, but we use this to demonstrate the desired effect of multi-component pulses. The dependence on the companion's relative position is set by
\begin{equation}
    f_\text{orbit}(\beta) = \exp \left( \frac{-\beta_\text{MD}^2}{2(W_\text{orbit}/5)^2} \right),
\end{equation}
where $W_\text{orbit}$ sets the angle from $\hat{\mu}$ where interaction with the companion activates. The resulting observed light curve is
\begin{equation}\label{eqn:flux}\begin{split}
    &I_\text{pred}(\beta,\beta_\text{MD}) =\\ &C [f_\text{spin}(\beta) f_\text{orb}(\beta_\text{MD}) + f_\text{spin}(\beta-\pi) f_\text{orb}(\beta_\text{MD}-\pi)]
\end{split}\end{equation}
where $C$ is a coefficient with units of Jy.

%%%%%%%%%% ALIASING ACTIVITY WINDOW %%%%%%%%%%
Here, it is assumed that the radio beam is centred on the white dwarf's magnetic axis and rotates with the spin, but the predicted light curve can be consistent with other arrangements, for example, the beam might be offset from the magnetic axis or not originate at the white dwarf's pole at all. In the case of a synchronised binary, the radio period should equal the orbital period. Meanwhile, if the spin and orbit are very different, like \var{1839}, then the radio period matches the spin period, with the intensity modulated on the orbital period. We call the range of orbital phases within which pulses are observed the \textit{orbital window}.

\begin{figure}[ht]
    \centering
    \includegraphics[width=\linewidth]{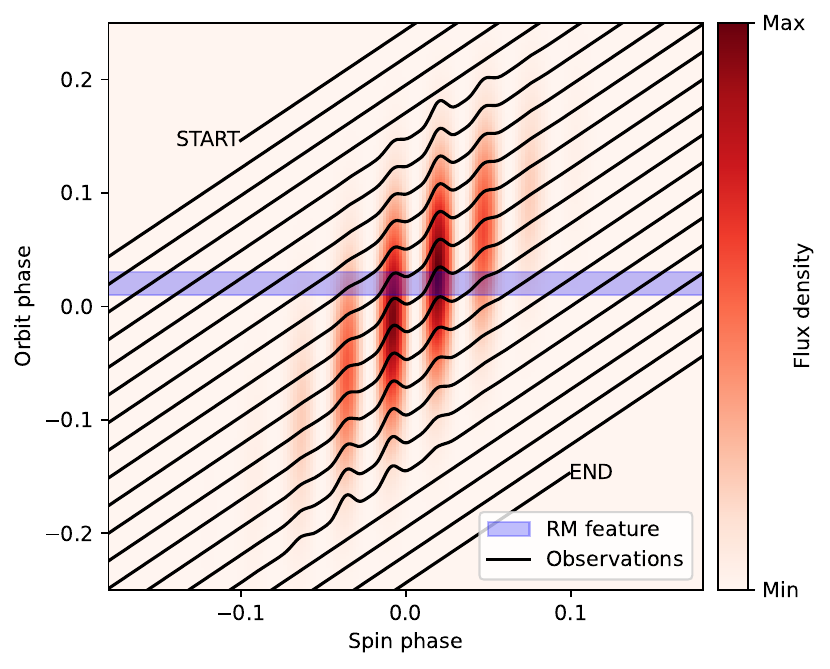}
    \caption{Proposed illustrative dynamic pulse profile of \var{1832} created using the geometric binary model described in \autoref{subsec:geometry} with the following parameters: $i=20^\circ$, $\alpha=45^\circ$, $\phi_0=15^\circ$, $\zeta=45^\circ$, $W_\text{spin}=30^\circ$, $W_\text{orb}=60^\circ$, $\theta_\text{struc}=0.16$. The background red shades represent the predicted flux density. The black lines represent the peak-normalised pulse profiles in a continuous observation, which are cross-sections through the dynamic pulse profile shown in red in the background. The cross-sections are chronologically top-to-bottom and have a gradient close to 1 due to the spin and orbit being nearly synchronised ($P_\text{orbit}=1.025 P_\text{spin}$). The horizontal blue line is a feature constant in orbital phase.}
    \label{fig:hotdog}
\end{figure}

However, if the orbital period is close to the spin period, then an aliasing situation can occur in which the apparent modulation period is \textit{much} longer than the orbital period. \autoref{fig:hotdog} shows the dynamic pulse profile (a map of observed intensity as a function of spin and orbital phase) for a toy system with $P_\text{orbit}=1.025P_\text{spin}$. Here, the white dwarf magnetic axis takes many orbits to become misaligned from the companion, hence creating an apparent activity window much longer than the orbit. The black lines represent a continuous observation, which have a gradient close to 1 due to the near-synchronicity. Although the orbital window is only a fraction of the orbit, the apparent window of activity is many orbits --- over 2 years in the case of \var{1832}.

In \autoref{fig:hotdog}, the model from \cite{2026NatAs.tmp...27H} is modulated by a function periodic in spin phase to simulate the substructure (\autoref{eqn:beam}), resulting in the vertical stripes. With nearly synchronised spin and orbital periods, and slightly misaligned spin and orbital axes, the pulse centre tends to drift towards negative spin phase while the pulse components stay fixed. This is a natural prediction of the model which phenomenologically agrees with our \var{1832} observations without requiring especially contrived parameters. Additionally, features constant in orbital phase drift towards positive pulse phase, as explained in the following section.

\subsection{Rotation Measure and spectral index}

%%%%%%%%%% RM / ALPHA CORRELATION %%%%%%%%%%
% P04--14 and the middles of P15--24.
RM and $\alpha_I$ are significantly correlated, peaking at the same pulse phases in P04--14 and P15--24, and staying constant in the rest of the pulses. Only P00--35 had high enough signal-to-noise to accurately measure the phase-resolved RM, but the rest of the pulses are included in \autoref{fig:1832_pulsestack} as well. Correlation between RM and $\alpha_I$ implies that the same effect is responsible for both.
Variation in RM and $\alpha_I$ is significantly reduced in P25--35, but this does not coincide with changes in other emission characteristics such as the PA curve or polarised fraction, implying that the RM / $\alpha_I$ feature in P04--14 and P15--24 are not intrinsic to the emission mechanism, but rather a propagation effect through a medium which enters and exits the line of sight. This medium must be (nearly) co-rotating with the emission site for the RM and $\alpha_I$ peaks to stay steady across consecutive pulses, and it must apply both a frequency-dependent rotation to the PA, as well as steepening the spectrum. \citep{1985ARA&A..23..169D, 2017RvMPP...1....5M}.

The core components of pulsar beams are expected to have steeper spectra than the surrounding cone components \citep{2022ApJ...927..208B}, but given that the RM / $\alpha_I$ feature in \var{1832} varies in pulse phase, we disfavour this interpretation. Mode-switching in pulsars is also often associated with changes in spectral index; some pulsars alternate between ``modes'' with distinct pulse profiles, spectral shapes, linear and circular polarised fractions, and position angle tracks \citep{1982ApJ...258..776B,2011ApJ...741...48C,2016Ap&SS.361..261W,2021MNRAS.502..127L}. Mode switches are thought to be triggered by changes in magnetospheric conditions, which might also affect the RM.
% It is peculiar, though, that the appearance of the central RM / $\alpha_I$ feature at P15 does not coincide with a significant change in the pulse profile.
Observations between P14 / P15 and P24 / P25 would have allowed us to see whether the RM / spectral index feature and profile change appears suddenly, supporting a neutron star interpretation, or drifts through pulse phase, supporting our model.

\cite{2026ApJ...997..124Y} explicitly predicts Faraday rotation in the magnetospheres of mCV-like LPTs, but the spectral steepening is more puzzling. 
Induced Compton scattering might down-shift photon frequencies \citep{1978MNRAS.185..297W, 1982MNRAS.200..881W}, absorption can alter the spectrum, or there may be chromatic dispersion at a boundary in plasma density \citep{1994ApJ...422..304T}, although steepening would be unusual in the latter two cases. A natural way to link RM and $\alpha_I$ is if the effect is mode-dependent --- spectral variation in just one orthogonal polarisation mode would change the overall spectrum, as well as contribute apparent RM. More detailed modelling would be required to understand this feature fully.

The medium imparting this effect on the beam may be a structure in the binary magnetosphere, such as a magnetically-channelled accretion stream (plausible given the detection of X-rays, see \autoref{subsec:xray}), or possibly the wind / magnetosphere of the companion (in a similar vein to McSweeney et al. 2026, submitted). In either case, as a structure locked to the binary frame, it would be fixed in orbital phase, represented by the blue horizontal region in \autoref{fig:hotdog}. Given that the emission mechanism is thought to be powered by a binary interaction, it makes sense that such a propagation effect would be observed when the source is brightest. Features locked in orbital phase on the dynamic pulse profile appear to drift towards positive spin phase (successive cross-sections cut the blue region at increasing spin phase in \autoref{fig:hotdog}), which is exactly what the RM / $\alpha_I$ feature appears to do between P06--14 and P15--24.

%%%%%%%%%% RM FEATURE DRIFT RATE %%%%%%%%%%
We can use the assumption that the RM / $\alpha_I$ feature is constant in orbital phase on the dynamic pulse profile to predict when \var{1832} will re-brighten.
There are 73 days between P24 and P25, over which time the RM / $\alpha_I$ feature must drift by at least 150\,s in spin phase to not be present in P25. This yields a minimum drift-rate of 2.05\,s/day. Over the 658 days between the first and last pulses, it would drift by $>1349$\,s, about half of \var{1832p1}. It would take $<1296$\,days or $<3.55$\,years to drift by a full spin period, so \var{1832} should re-brighten by July of 2027.
The fact that re-brightening did not occur in 658 days gives a maximum drift-rate of $3.51$\,s/day. In the 2.7 days between P14 and P15 the RM / $\alpha_I$ feature would drift by 5.5\,s to 9.5\,s, which appears consistent with observation.

\subsection{X-rays}\label{subsec:xray}

%%%%%%%%%% X-RAY DIP FEATURE %%%%%%%%%%
\cite{2025Natur.642..583W} observed pulsed X-ray emission from \var{1832} at the radio period, as well as a dip in X-ray luminosity to 0 photons at $\sim0.1-0.3$ pulse phase (see \citeauthor{2025Natur.642..583W}'s Figure. 3). They argue that the X-ray dip cannot be explained as occlusion by a companion because the companion would have to be too big or too close, but if occlusion is by a structure in the interacting magnetosphere like an accretion stream channelled by the white dwarf magnetic field, or the white dwarf itself, then this feature can make sense. The time of this observation is shown in red in the right-most panel of \autoref{fig:1832_pulsestack}. \var{1832} was in its brightest radio state at the time, and by the time of follow-up X-ray observations which made no detection, the radio brightness had decreased by 3 orders of magnitude. \citeauthor{2025Natur.642..583W} concluded that the source has a constant ratio of X-ray to radio luminosity.
In any case, the presence of X-rays in the binary interpretation, and especially the coincidence of the radio and X-ray periods, is suggestive of accretion onto the white dwarf powering the emission at both wavelengths.

% Paragraph from Nanda:
An interesting characteristic of asynchronous polars is their tendency to exhibit exceptionally large X-ray pulsed fractions, with pulse profiles that can decrease to nearly zero flux at minimum. Such deep modulations are uncommon among other accreting white dwarf systems, particularly intermediate polars, where emission from extended accretion curtains and multiple accreting regions generally prevents complete disappearance of the X-ray signal. In asynchronous polars, however, accretion is often concentrated onto a compact magnetic pole, allowing geometric self-occultation by the white dwarf and, in some cases, additional absorption by the accretion stream to produce extremely deep minima \citep{1990SSRv...54..195C, 2002MNRAS.335..918R, 2017PASP..129f2001M}. The slight mismatch between the spin and orbital periods further modifies the accretion geometry over the beat cycle, potentially enhancing the visibility changes of the dominant accretion region. The high X-ray pulsed fraction observed here is therefore consistent with the strongly magnetic and highly localised accretion flow characteristic of asynchronous polars. Gradual changes in spectral features and duty cycle associated with the asynchronism should be detectable if the source re-brightens.

\subsection{Linear position angle and circular polarisation}

In the canonical picture of pulsar radio emission, the PA sweeps an ``S'' shape through pulse phase as the field lines of the dipolar magnetic field pass the line of sight \citep{1969ApL.....3..225R}. This is known as the rotating vector model (RVM). While the PA sweep is not a unifying feature of pulsars, deviations from the RVM can often be attributed to a combination of orthogonal polarisation modes \citep{2023MNRAS.525..840O} and scattering by the interstellar medium \citep{2009MNRAS.392L..60K}. \var{1832} is not significantly scattered (evidenced by the high polarisation fraction and the long duration of the pulses compared with typical scattering timescales), but the partially-coherent superposition of orthogonal linear polarisation modes at various ratios naturally predicts linear-to-circular rotation of the Stokes vector \citep{2023MNRAS.525..840O}, explaining the coincidence of PA sweeps and increased circular fraction in \autoref{fig:1832_pa_vfrac_phase}.

\var{1839} displays two orthogonal PA modes which depends on the location in the dynamic pulse profile \citep{2026NatAs.tmp...27H}, which was interpreted as a birefringent medium refracting orthogonal modes. The presence of birefringence is even more evident in ASKAP\,J142431.2\ensuremath{-}612611 \citep[][accepted]{2026arXiv260307857P}. In the case of \var{1832}, we may be more slowly probing regions of the dynamic pulse profile where one or another PA mode is dominant, with increased circular polarisation at the boundaries of those regions where modes superimpose. This is especially attractive given the PA curve seems to evolve on the same slow timescale as the pulse morphology (\autoref{fig:1832_pulsestack}).

PA shapes changing from pulse to pulse have been seen in magnetars like \var{1810} \citep{2024NatAs...8..617D, 2024NatAs...8..606L}.
The PA sweep of this magnetar drifts in pulse phase and flips in sign, somewhat resembling the first and last epochs of \autoref{fig:1832_pa_vfrac_phase}. ``Off-center'' RVM PA sweeps and vertical offsets are an expected phenomenon in magnetar magnetic fields \citep{2009ApJ...703.1044B,2002ApJ...574..332T,2021MNRAS.502.1549T,2017MNRAS.466L..73P}, and sign flips can be achieved with precession changing the viewing angle. Linear-to-circular conversion and fluctuations in the magnetosphere complicate matters further. These effects may play a role in \var{1832} and other LPTs as well, whether they are of white dwarf or magnetar origin, and further work will be needed to build a comprehensive picture.

\section{Conclusion}

The evolution of the pulse morphology and the rotation measure variation of \var{1832 long} can be explained in a nearly synchronised close binary model. The quasi-periodicity, PA, circular polarisation, X-ray emission, and spectral index are, at least, consistent with this interpretation as well. In addition, we make the testable prediction of re-brightening by July of 2027, allowing it to be strengthened or weakened in the future.

This work does not definitively identify \var{1832} as a white dwarf binary, and none of the observational properties of \var{1832} definitively excludes the magnetar (or other) interpretation.
Slow pulse profile evolution, phase-resolved RM and spectral index variation, and non-RVM-like PA curves are frequently seen among pulsars and magnetars. Higher cadence observations while the source was bright would have enabled a more thorough test of our model, such as whether the RM / $\alpha_I$ feature drifts or appears suddenly.
However, a synthesis of all of the observed properties, as well as the building body of knowledge about other LPTs, leads us to prefer an ultra-compact asynchronous polar interpretation.

We note that a number of other shorter-period $P\lesssim1.5$\,hr LPTs with yet-unknown progenitors have similar phenomenology to \var{1832}: slow pulse profile evolution with a transient $\sim$weeks--months window of activity. These include \var{1627} \citep{2022Natur.601..526H}, ASKAP\,J142431.2\ensuremath{-}612611 \citep[][accepted]{2026arXiv260307857P}, ASKAP\,J175534.9$-$252749.1 \citep{2025MNRAS.542..203M}, and potentially even the original Galactic ``burper'' \citep{2005Natur.434...50H}. Near-synchronous binaries may explain these phenomena.  If so, they may be detectable by proposed space-based gravitational wave detectors like the Laser Interferometer Space Antenna \citep[LISA;][]{2017arXiv170200786A} and Taiji \citep{2020ResPh..1602918L}, according to \cite{2025ApJ...991..134S}. Longer-period sources may potentially re-brighten, whereas shorter-period sources may be more likely to be disrupted by tidal effects, and even undergo mergers. A larger sample of LPTs with these particular radio characteristics, clear evidence of periodic re-brightening, and/or identification of suitable optical counterparts, would help test this hypothesis. Large radio interferometric surveys and archival searches are currently underway across multiple observatories and will likely provide the required number statistics in coming years.

\appendix

\section{Dynamic spectrum extraction}

The calibrated ASKAP measurement sets for each PAF beam with sensitivity to \var{1832} were downloaded from the CSIRO ASKAP Science Data Archive  \footnote{https://data.csiro.au/domain/casdaObservation/}. They were re-phased to reference the centre of each beam (as opposed to the centre of the nominal pointing of the entire PAF footprint), and imaged with \textsc{WSClean} \citep{2014MNRAS.444..606O} to determine a field source model. During this process, a mask was placed at the location of \var{1832} to prevent it from being \textsc{clean}ed into the model. The field model was subtracted from the visibilities, and the data re-phased to the location of \var{1832}. Baselines longer than 100\,m were averaged (equivalent to forming a naturally-weighted beam on the source) to produce a dynamic spectrum in time, frequency, and instrumental polarisation. The instrumental polarisations were converted to celestial Stokes by the standard transforms and corrected for the primary beam.
Where multiple beams had sensitivity to the same pulse, the dynamic spectra were averaged together, weighted by the primary beams to produce a single dynamic spectrum.
Parallactic angle correction did not need to be applied due to ASKAP's roll-mount design. Finally, the dynamic spectra were de-dispersed to $\rm{DM}=458$\,pc\,cm$^{-3}$ \citep{2025Natur.642..583W}. The results had 288 channels (except for the 1296--1439\,MHz band with 144 channels) and a 2\,s time integration.

\section{Dynamic spectrum model}\label{subsec:dynspec_model}
\renewcommand{\thefigure}{B\arabic{figure}} % Formats the label as A1, A2, etc.
\setcounter{figure}{0}  

The spectra of Stokes I and V are modelled as power laws for every time-bin $t$ according to
\begin{equation}
    I'_t(\nu) = I_{\text{1GHz}\,t} \left(\frac{\nu}{\text{1GHz}}\right)^{\alpha_{I\,t}}
    \label{eqn:stokesI}
\end{equation}
\begin{equation}
    \text{and}~~~ V'_t(\nu) = V_{\text{1GHz}\,t} \left(\frac{\nu}{\text{1GHz}}\right)^{\alpha_{V\,t}}
    \label{eqn:stokesV}
\end{equation}
where $\nu$ is frequency, $I_{\text{1GHz}\,t}$ and $V_{\text{1GHz}\,t}$ are the modelled I and V at 1GHz, and $\alpha_{I\,t}$ and $\alpha_{V\,t}$ are the spectral indices of I and V respectively.

The linear polarised component, Stokes Q and U, is modelled for every time-bin according to
\begin{equation}
    \left[\begin{matrix}Q'_t(\nu) \\ U'_t(\nu)\end{matrix}\right] =
    L_{\text{1GHz}\,t} \left(\frac{\nu}{\text{1GHz}}\right)^{\alpha_{L\,t}}
    \left[\begin{matrix}\cos \psi_t(\nu) \\ \sin \psi_t(\nu) \end{matrix}\right]
    \label{eqn:stokesL}
\end{equation}
\begin{equation}
    \text{where}~~~ \psi_t(\nu) = 2 \left( PA_t + RM_t \left(\frac{c}{\nu}\right)^2 \right)
    \label{eqn:psi}
\end{equation}
is the polarisation position angle which has been Faraday rotated by propagation through the magnetised interstellar medium, $Q'_t(\nu)$ and $U'_t(\nu)$ are the modelled Q and U spectra, $L_{\text{1GHz}\,t}$ is the modelled L at 1GHz, $\alpha_{L\,t}$ is the spectral index of L, $PA_t$ is the polarisation position angle at the source, $RM_t$ is the Faraday rotation measure, and $c$ is the speed of light in vacuum.

\begin{figure}[t]
    \centering
    \includegraphics[width=\linewidth]{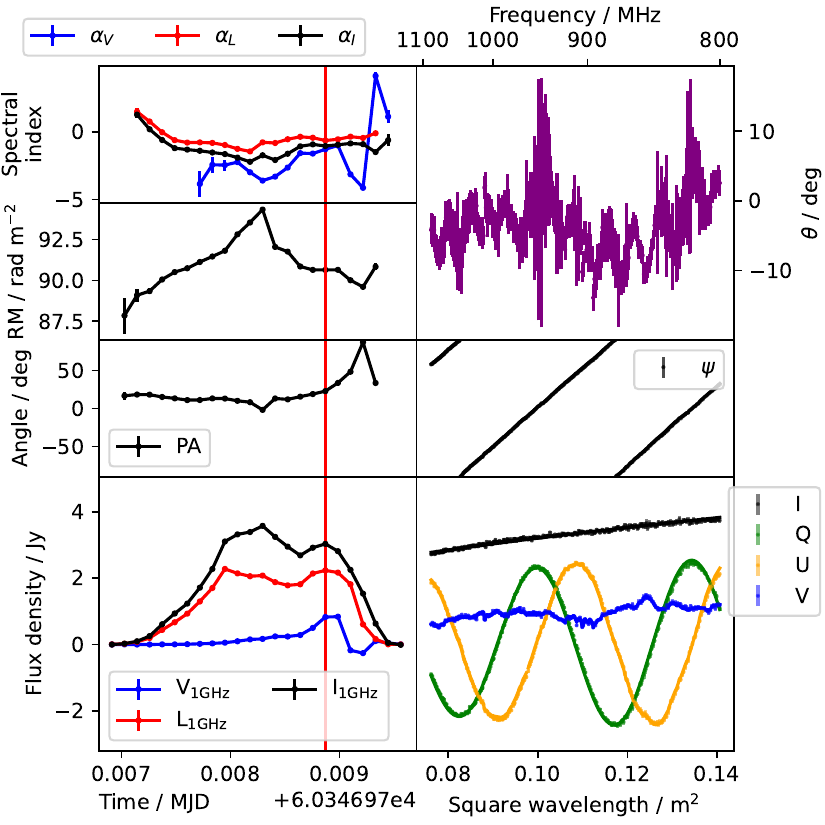}
    \caption{Measured properties of P18 from SB58609. At left are time-dependent values. At right are frequency-dependent values at the time marked by the red vertical line (except for $\theta$ which is the same for the whole observation). At right, points with error bar mark the measured values in the dynamic spectrum, and continuous curves mark the modelled I, Q, U, V, and PA. The curves mostly overlap the points, indicating good agreement between model and measurement.}
    \label{fig:rep_pulse}
\end{figure}

The brightest pulses (P04--35) had frequency-dependent circular polarisation features in their dynamic spectra (\autoref{fig:rep_pulse} shows a representative pulse) which might, at first glance, be misinterpreted as an intrinsic feature such as generalised Faraday rotation. However, we show that it is actually leakage between Stokes U and Stokes V made visible by the very high signal to noise ratio. When fractional V is large, the frequency-dependent structure transfers to the U spectrum, and only ever the U spectrum. Furthermore, the angle of U to V rotation is constant across entire observations with many pulses and even between observations. The leakage is not a result of incorrect primary beam correction, because such rapid spectral variation is uncharacteristic of the primary beam, which varies smoothly.  Note that the P04--35 data are from the same ASKAP beam.
We model the leakage as a rotation in Stokes space in the Stokes (U,V) plane by a time-independent, frequency-dependent angle $\theta_\nu$ according to
\begin{equation}
    \left[\begin{matrix}U''_{t,\nu} \\ V''_{t,\nu}\end{matrix}\right] =
    \left[\begin{matrix}\cos\theta_\nu & -\sin\theta_\nu \\ \sin\theta_\nu & \cos\theta_\nu\end{matrix}\right]
    \left[\begin{matrix}U'_t(\nu)\\ V'_t(\nu)\end{matrix}\right]
    \label{eqn:leakage_model}
\end{equation}
where $U''_{t,\nu}$ and $V''_{t,\nu}$ are the leakage-modelled U and V at frequency $\nu$. The measured $U_{t,\nu}$ and $V_{t,\nu}$ can be corrected for this leakage by rotating by $\theta_\nu$ in the opposite direction as in
\begin{equation}
    \left[\begin{matrix}U^*_{t,\nu} \\ V^*_{t,\nu}\end{matrix}\right] =
    \left[\begin{matrix}\cos\theta_\nu & \sin\theta_\nu \\ -\sin\theta_\nu & \cos\theta_\nu\end{matrix}\right]
    \left[\begin{matrix}U_{t,\nu}\\ V_{t,\nu}\end{matrix}\right]
    \label{eqn:leakage_correction}
\end{equation}
where $U^*_{t,\nu}$ and $V^*_{t,\nu}$ are the leakage-corrected U and V.

All fits were done per de-dispersed dynamic spectrum time-step using least-squares regression with \verb|scipy.optimize.curve_fit|, setting the \verb|sigma| uncertainty argument to the per-channel RMS noise $\sigma_\nu$. We measure $\theta_\nu$ using the following procedure:
\begin{enumerate}
    \item Initialise $\theta_\nu := 0$.
    \item Calculate $(U^*_{t,\nu}, V^*_{t,\nu})$ using \autoref{eqn:leakage_correction}.
    \item Find optimal $V_{\text{1GHz}\,t}$ and $\alpha_{V\,t}$ by fitting $V'_t(\nu)$ to $V^*_{t,\nu}$ using \autoref{eqn:stokesV}.
    \item Find optimal $L_{\text{1GHz}\,t}$, $\alpha_{L\,t}$, $RM_t$, and $PA_t$ by fitting $(Q'_t(\nu), U'_t(\nu))$ to $(Q_{t,\nu}, U^*_{t,\nu})$ using \autoref{eqn:stokesL}.
    \item Estimate $\theta_\nu$ as the weighted time-mean angle from $(U'_t(\nu), V'_t(\nu))$ to the original $(U_{t,\nu}, V_{t,\nu})$. To avoid the angle-wrapping problem, this is done in complex space defining $R'_t(\nu) = U'_t(\nu) + i V'_t(\nu)$ and $R_{t,\nu} = U_{t,\nu} + i V_{t,\nu}$ where $i$ is the imaginary unit. Then,
    \begin{equation}
        \theta_\nu = \arg\left\{\frac{1}{\sum_t W_{t,\nu}} \sum_t \left[ W_{t,\nu} \frac{R_{t,\nu}}{R'_t(\nu)} \frac{|R'_t(\nu)|}{|R_{t,\nu}|} \right]\right\}
    \end{equation}
    where $W_{t,\nu}$ is the weighting with low signal-to-noise data weighted to zero as in
    \begin{equation}
    W_{t,\nu} = \left\{\begin{matrix}
        |R_{t,\nu}| ~/~ \sigma_\nu &\text{if } |R_{t,\nu}| > 5\sigma_\nu;\\
        0 & \text{otherwise.}
    \end{matrix}\right.\end{equation}
    \item Repeat twice more from step 2 for an improved estimate of $\theta_\nu$.
\end{enumerate}

\section{Calculations}
\renewcommand{\thefigure}{C\arabic{figure}} % Formats the label as A.1, A.2, etc.
\setcounter{figure}{0}  

\begin{figure}[h]
    \centering
    \includegraphics[width=\linewidth]{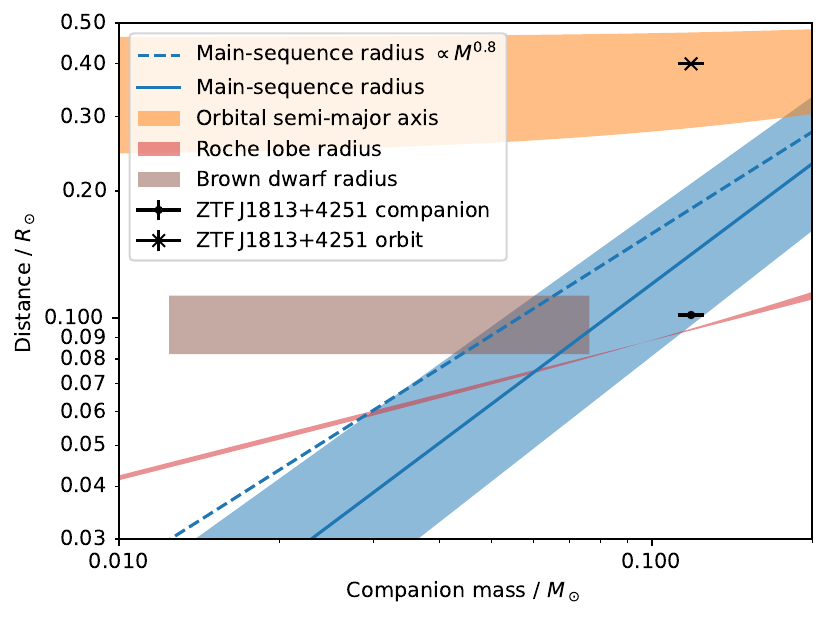}
    \caption{Possible mass, radius, and orbital separation of the companion of a white dwarf in \var{1832}. The radius of a main-sequence companion is shown using \autoref{eqn:simple_radius} and \autoref{eqn:empirical_radius}. The orbital semi-major axis (\autoref{eqn:semi-major_axis}) and Roche-lobe radius (\autoref{eqn:roche_lobe}) are calculated assuming $P_\text{orbit} = 44.3$\,min and $M_s \in [0.2, 1.4]\,\rm{M}_\odot$. The radius and mass range of a brown dwarf companion is included. The cataclysmic variable \var{1813} with a 51\,min orbit is included for reference for which \cite{2022Natur.610..467B} found remarkably accurate mass measurements.}
    \label{fig:dist_mass}
\end{figure}

\autoref{fig:dist_mass} shows the possible orbital semi-major axis, Roche lobe radius, and radius of the white dwarf's companion if it were main-sequence. \var{1813} with a 51\,min orbit \cite{2022Natur.610..467B} is included as a concrete example of parameters for such a tight cataclysmic variable. The orbital semi-major axis was calculated using
\begin{equation}\label{eqn:semi-major_axis}
    a^3 = \frac{G(M_s+M_c)P_\text{orbit}^2}{4\pi^2}.
\end{equation}
where $P_\text{orbit}$ is the orbital period and $G$ is the gravitational constant. The Roche lobe radius is approximated as
\begin{equation}\label{eqn:roche_lobe}
    \frac{R_{rl}}{a} = \frac{0.49 q^{2/3}}{(0.6q^{2/3} + \ln(1+q^{1/3}))}
\end{equation}
where $q = M_c / M_s$ \citep{1983ApJ...268..368E}. The main-sequence radius $R_c$ is shown in \autoref{fig:dist_mass} using both
\begin{equation}\label{eqn:simple_radius}
    R_c = (M_c / \rm{M_\odot})^{0.8} \rm{R}_\odot
\end{equation}
and
\begin{equation}\label{eqn:empirical_radius}
    \log_{10} (R_c / \rm{R}_{\odot}) = (0.026\pm0.13)+(0.945\pm0.041)\log_{10} (M_c / \rm{M}_{\odot})
\end{equation}
found empirically by \cite{1991Ap&SS.181..313D}. The in-spiral timescale due to gravitational waves is estimated using
\begin{equation}
    \tau_{\rm{GW}} = \left(3.3\times10^{17}\text{ years}\right) \left(\frac{a}{\text{AU}}\right)^4 \left(\frac{M_s M_c (M_s + M_c)}{\rm{M}_\odot^3}\right)^{-1}
\end{equation}
\citep{1963PhRv..131..435P}. For a white dwarf of $M_s=1\,\rm{M}_\odot$ with a companion of $M_c=0.1\,\rm{M}_\odot$, we find $\tau_{\rm{GW}}\approx48$\,M\,years.

\section{Luminosity predictions} \label{sec:luminosity}

We consider a white dwarf of mass $M_s \approx 1\rm{~M_\odot}$, radius $R_s \approx 0.0075\rm{~R_\odot}$, and surface magnetic field $B_s \in [10^6, 10^9]\rm{~G}$ with a companion of mass $M_c \approx 0.1\rm{~M_\odot}$. The white dwarf spin period is $P_s = 2656.25325 \rm{~s}$ and the orbit is $P=(1+P_s/3.5\rm{~yrs}) P_s = 2656.317130 \rm{~s}$. The radio luminosity predicted by \citeauthor{2025ApJ...981...34Q}'s Equation 11 is
\begin{equation}\begin{aligned}
    \dot{E}_{\text{\citeauthor{2025ApJ...981...34Q}}}
    \approx& (2\times10^{25}\rm{~erg~s^{-1}}) \zeta^2 \left(\frac{B_s}{1\times10^6\rm{~G}}\right)^2 \left(\frac{R_c}{0.2\rm{~R_\odot}}\right)^2 \\ & \left(\frac{M_s}{0.8 \rm{~M_\odot}}\right)^{-4/3}\left(\frac{P}{125.5\rm{~min}}\right)^{-14/3} \\
    \in& [7\times10^{17}, 7\times10^{23}]
\end{aligned}\label{eqn:qu_zhang}\end{equation}
where $\zeta=|1-\Omega_s/\Omega|$, $\Omega_s=2\pi/P_s$, and $\Omega=2\pi/P$. The radio luminosity predicted by \citeauthor{2026ApJ...997..124Y}'s Equation 32 is
\begin{equation}\begin{aligned}
    \dot{E}_{\text{\citeauthor{2026ApJ...997..124Y}}}
    \approx& (3.9\times10^{28}\rm{~erg~s^{-1}}) \left(\frac{\mu_s}{1\times10^{34}\rm{~G~cm^3}}\right)^2\left(\frac{R_c}{1\times10^{10}\rm{~cm}}\right)^2\\& \left(\frac{P}{100\rm{~min}}\right)^{-14/3} \left(\frac{M_s+M_c}{\rm{~M_\odot}}\right)^{-4/3} \left(\frac{\Delta\Omega}{\Omega}\right)^2 \\
    \in& [2\times10^{17}, 2\times10^{23}]
\end{aligned}\label{eqn:yang}\end{equation}
where $\mu_s=B_s R_s^3$ and $\zeta_\phi = 4 \Delta\Omega a / (\pi c)$. The maximum radio luminosity predicted by \citeauthor{2026ApJ...999L...2Z}'s Equation 1 is
\begin{equation}\begin{aligned}
    \dot{E}_{\text{\citeauthor{2026ApJ...999L...2Z}}}
    \approx& (6\times10^{29}\rm{~erg~s^{-1}}) \zeta_\phi \left(\frac{\Delta\Omega}{\Omega}\right) \left(\frac{\mu_s}{1\times10^{33}\rm{~G~cm^3}}\right)^2\\& \left(\frac{R_c}{0.217\rm{~R_\odot}}\right)^2\left(\frac{M_s+M_c}{0.8\rm{~M_\odot}}\right)^{-5/3} \left(\frac{P}{100\rm{~min}}\right)^{-13/3} \\
    \in& [2\times10^{17}, 2\times10^{23}]
\end{aligned}\label{eqn:zhong}\end{equation}
All of these are much less than the $\sim10^{32}$\,erg~s$^{-1}$ \citep{2025Natur.642..583W} observed in the brightest phase of \var{1832}.

\paragraph{Acknowledgements}
This scientific work uses data obtained from Inyarrimanha Ilgari Bundara, the CSIRO Murchison Radio-astronomy Observatory. We acknowledge the Wajarri Yamaji People as the Traditional Owners and native title holders of the Observatory site. CSIRO’s ASKAP radio telescope is part of the Australia Telescope National Facility (\url{https://ror.org/05qajvd42}). Operation of ASKAP is funded by the Australian Government with support from the National Collaborative Research Infrastructure Strategy. ASKAP uses the resources of the Pawsey Supercomputing Research Centre. Establishment of ASKAP, Inyarrimanha Ilgari Bundara, the CSIRO Murchison Radio-astronomy Observatory and the Pawsey Supercomputing Research Centre are initiatives of the Australian Government, with support from the Government of Western Australia and the Science and Industry Endowment Fund.

We thank Prof. Nanda Rea for useful feedback on the interpretation of the X-ray characteristics of \var{1832}.

\paragraph{Funding statement}
This research is supported by an Australian Government Research Training Program (RTP) Scholarship {doi.org/10.82133/C42F-K220}.
Ziteng Wang is funded by an Australian Research Council Discovery Project (project number FT190100231).

% \paragraph{Competing Interests}
% The authors declare no competing interests.

\paragraph{Data availability statement}
The dynamic spectra and scripts used to process them are available at \url{https://github.com/CsanadHorvath/J1832-0911_binary}.

%%%%%%%%%%%%% JOURNAL ABREVIATIONS %%%%%%%%%%%%%
\newcommand{\actaa}{Acta Astronomica}
\newcommand{\araa}{Annual Review of Astron and Astrophys}
\newcommand{\areps}{Annual Review of Earth and Planetary Science}
\newcommand{\aar}{Astrononmy and Astrophysics Review}
\newcommand{\ab}{Astrobiology}
\newcommand{\aj}{Astronomical Journal}
\newcommand{\ac}{Astronomy and Computing}
\newcommand{\apart}{Astroparticle Physics}
\newcommand{\apj}{Astrophysical Journal}
\newcommand{\apjl}{Astrophysical Journal, Letters}
\newcommand{\apjs}{Astrophysical Journal, Supplement}
\newcommand{\ao}{Applied Optics}
\newcommand{\apss}{Astrophysics and Space Science}
\newcommand{\aap}{Astronomy and Astrophysics}
\newcommand{\aapr}{Astronomy and Astrophysics Reviews}
\newcommand{\aaps}{Astronomy and Astrophysics, Supplement}
\newcommand{\baas}{Bulletin of the AAS}
\newcommand{\caa}{Chinese Astronomy and Astrophysics}
\newcommand{\cjaa}{Chinese Journal of Astronomy and Astrophysics (now RAA)}
\newcommand{\cqg}{Classical and Quantum Gravity}
\newcommand{\epsl}{Earth and Planetary Science Letters}
\newcommand{\expa}{Experimental Astronomy}
\newcommand{\frass}{Frontiers in Astronomy and Space Sciences}
\newcommand{\gal}{Galaxies}
\newcommand{\gca}{Geochimica Cosmochimica Acta}
\newcommand{\grl}{Geophysics Research Letters}
\newcommand{\icarus}{Icarus}
\newcommand{\ija}{International Journal of Astrobiology}
\newcommand{\jatis}{Journal of Astronomical Telescopes, Instruments, and Systems }
\newcommand{\jcap}{Journal of Cosmology and Astroparticle Physics}
\newcommand{\jgr}{Journal of Geophysics Research}
\newcommand{\jgrp}{Journal of Geophysics Research: Planets}
\newcommand{\jheap}{Journal of High Energy Astrophysics}
\newcommand{\joss}{Journal of Open Source Software}
\newcommand{\jqsrt}{Journal of Quantitiative Spectroscopy and Radiative Transfer}
\newcommand{\lrca}{Living Reviews in Computational Astrophysics}
\newcommand{\lrr}{Living Reviews in Relativity}
\newcommand{\lrsp}{Living Reviews in Solar Physics}
\newcommand{\memsai}{Mem. Societa Astronomica Italiana}
\newcommand{\maps}{Meteoritics and Planetary Science}
\newcommand{\mnras}{Monthly Notices of the RAS}
\newcommand{\nat}{Nature}
\newcommand{\nastro}{Nature Astronomy}
\newcommand{\ncomms}{Nature Communications}
\newcommand{\ngeo}{Nature Geoscience}
\newcommand{\nphys}{Nature Physics}
\newcommand{\na}{New Astronomy}
\newcommand{\nar}{New Astronomy Review}
\newcommand{\physrep}{Physics Reports}
\newcommand{\pra}{Physical Review A: General Physics}
\newcommand{\prb}{Physical Review B: Solid State}
\newcommand{\prc}{Physical Review C}
\newcommand{\prd}{Physical Review D}
\newcommand{\pre}{Physical Review E}
\newcommand{\prl}{Physical Review Letters}
\newcommand{\psj}{Planetary Science Journal}
\newcommand{\planss}{Planetary Space Science}
\newcommand{\pnas}{Proceedings of the US National Academy of Sciences}
\newcommand{\procspie}{Proceedings of the SPIE}
\newcommand{\pasa}{Publications of the Astron. Soc. of Australia}
\newcommand{\pasj}{Publications of the Astron. Soc. of Japan (note no full stop following Jpn)}
\newcommand{\pasp}{Publications of the Astron. Soc. of the Pacific}
\newcommand{\raa}{Research in Astronomy and Astrophysics (formerly CJAA)}
\newcommand{\rasti}{RAS Techniques and Instruments}
\newcommand{\rmxaa}{Revista Mexicana de Astronomia y Astrofisica}
\newcommand{\rnaas}{Research Notes of the AAS.}
\newcommand{\sci}{Science}
\newcommand{\sciadv}{Science Advances}
\newcommand{\solphys}{Solar Physics}
\newcommand{\sovast}{Soviet Astronomy}
\newcommand{\ssr}{Space Science Reviews}
\newcommand{\uni}{Universe}
\newcommand{\aplett}{Astrophysics Letters}
\bibliographystyle{apj}
% \newpage
\bibliography{refs}

\end{document}